\documentclass[12pt]{iopart}
\usepackage{cite}

\usepackage{axodraw}  

\renewcommand{\lsim}
{\mathrel{\raisebox{-.3em}{$\stackrel{\displaystyle <}{\sim}$}}}

\def\asymp#1%
{\mathrel{\raisebox{-.4em}{$\widetilde{\scriptstyle #1}$}}}

\def\Nequal#1%
{\mathrel{\raisebox{-.5em}{$\stackrel{=}{\scriptstyle\rm#1}$}}}
\newcommand{\dsl}[1]{\not \hspace{-0.7mm}#1}
\def\dsl{\mathpalette\make@slash}
\def\make@slash#1#2{\setbox\z@\hbox{$#1#2$}%
  \hbox to 0pt{\hss$#1/$\hss\kern-\wd0}\box0}

\def\beq{\begin{equation}}
\def\eeq{\end{equation}}
\def\beqar{\begin{eqnarray}}
\def\eeqar{\end{eqnarray}}
\def\barr#1{\begin{array}{#1}}
\def\earr{\end{array}}
\def\bfi{\begin{figure}}
\def\efi{\end{figure}}
\def\btab{\begin{table}}
\def\etab{\end{table}}
\def\bce{\begin{center}}
\def\ece{\end{center}}
\def\nn{\nonumber}

\def\text{\textstyle}

\def\ga{\gamma}
\def\de{\delta}

\def\si{\sigma}

\def\reffi#1{\mbox{Figure~\ref{#1}}}

\def\refse#1{\mbox{Section~\ref{#1}}}

\def\citere#1{\mbox{Ref.~\cite{#1}}}
\def\citeres#1{\mbox{Refs.~\cite{#1}}}

\newcommand{\TeV}{\unskip\,\mathrm{TeV}}
\newcommand{\GeV}{\unskip\,\mathrm{GeV}}
\newcommand{\MeV}{\unskip\,\mathrm{MeV}}

\newcommand{\fb}{\unskip\,\mathrm{fb}}

\newcommand{\ri}{{\mathrm{i}}}
\newcommand{\rd}{{\mathrm{d}}}

\def\mathswitchr#1{\relax\ifmmode{\mathrm{#1}}\else$\mathrm{#1}$\fi}

\newcommand{\PW}{\mathswitchr W}

\newcommand{\PZ}{\mathswitchr Z}

\newcommand{\Pg}{\mathswitchr g}

\newcommand{\Pe}{\mathswitchr e}

\newcommand{\Pnebar}{\mathswitch \bar\nu_{\mathrm{e}}}
\newcommand{\Pd}{\mathswitchr d}
\newcommand{\Pdbar}{\bar{\mathswitchr d}}
\newcommand{\Pu}{\mathswitchr u}
\newcommand{\Pubar}{\bar{\mathswitchr u}}

\newcommand{\Pep}{\mathswitchr {e^+}}
\newcommand{\Pem}{\mathswitchr {e^-}}

\def\mathswitch#1{\relax\ifmmode#1\else$#1$\fi}

\newcommand{\MW}{\mathswitch {M_\PW}}

\newcommand{\Me}{\mathswitch {m_\Pe}}

\newcommand{\GW}{\Gamma_{\PW}}

\newcommand{\eeWWffff}{\Pep\Pem\to\PW\PW\to 4f}

\def\draftdate{\relax}
\def\mda{\relax}
\def\mua{\relax}
\def\mla{\relax}
\def\draft{
\def\thtystars{******************************}
\def\sixtystars{\thtystars\thtystars}
\typeout{}
\typeout{\sixtystars**}
\typeout{* Draft mode!
         For final version remove \protect\draft\space in source file *}
\typeout{\sixtystars**}
\typeout{}
\def\draftdate{\today}
\def\mua{\marginpar[\boldmath\hfil$\uparrow$]%
                   {\boldmath$\uparrow$\hfil}%
                    \typeout{marginpar: $\uparrow$}\ignorespaces}
\def\mda{\marginpar[\boldmath\hfil$\downarrow$]%
                   {\boldmath$\downarrow$\hfil}%
                    \typeout{marginpar: $\downarrow$}\ignorespaces}
\def\mla{\marginpar[\boldmath\hfil$\rightarrow$]%
                   {\boldmath$\leftarrow $\hfil}%
                    \typeout{marginpar: $\leftrightarrow$}\ignorespaces}
\def\Mua{\marginpar[\boldmath\hfil$\Uparrow$]%
                   {\boldmath$\Uparrow$\hfil}%
                    \typeout{marginpar: $\uparrow$}\ignorespaces}
\def\Mda{\marginpar[\boldmath\hfil$\Downarrow$]%
                   {\boldmath$\Downarrow$\hfil}%
                    \typeout{marginpar: $\downarrow$}\ignorespaces}
\def\Mla{\marginpar[\boldmath\hfil$\Rightarrow$]%
                   {\boldmath$\Leftarrow $\hfil}%
                    \typeout{marginpar: $\leftrightarrow$}\ignorespaces}
\overfullrule 5pt
\oddsidemargin -15mm
\marginparwidth 29mm
}

\begin{document}

\thispagestyle{empty}
\def\thefootnote{\fnsymbol{footnote}}
\setcounter{footnote}{0}
\null
\hfill BI-TP 99/46 \\
\strut\hfill ER/40685/940 \\
\strut\hfill LU-ITP 1999/021\\
\strut\hfill PSI-PR-99-30\\
\strut\hfill UR-1592\\
\strut\hfill hep-ph/9912290
\vfill
\begin{center}
{\Large \bf\boldmath
{Four-fermion production with {\sc RacoonWW}%
\footnote{To appear in the proceedings of the {\it UK Phenomenology
Workshop on Collider Physics}, Durham, UK, 19--24 September, 1999.}}
\par} 
\vspace{1cm}

{\large
{\sc A.\ Denner$^1$, S.\ Dittmaier$^2$, M. Roth$^{3}$ and 
D.~Wackeroth$^4$} } \\[1cm]

$^1$ {\it Paul Scherrer Institut\\
CH--5232 Villigen PSI, Switzerland} \\[0.5cm]

$^2$ {\it Theoretische Physik, Universit\"at Bielefeld \\
D--33615 Bielefeld, Germany}
\\[0.5cm]

$^3$ {\it Institut f\"ur Theoretische Physik, Universit\"at Leipzig\\
D--04109 Leipzig, Germany}
\\[0.5cm]

$^4$ {\it Department of Physics and Astronomy, University of Rochester\\
Rochester, NY 14627-0171, USA}
\par \vskip 1em
\end{center}\par
\vskip 2cm {\bf Abstract:} \par {\sc RacoonWW} is an event generator
for $\Pep\Pem\to\PW\PW\to 4\,$fermions$(+\gamma)$ that includes full
tree-level predictions for $\Pep\Pem\to 4f$ and $\Pep\Pem\to
4f+\gamma$ as well as ${\cal O}(\alpha)$ corrections to $\Pep\Pem\to4f$ 
in the so-called double-pole approximation. We briefly sketch the concept 
of the calculation on which this generator is based and present
some numerical results.
\par
\vskip 1cm
\noindent
December 1999

\null
\setcounter{page}{0}

\clearpage

\title[Four-fermion production with {\sc RacoonWW}]%
{Four-fermion production with {\sc RacoonWW}}

\author{A.\ Denner$^1$, S.\ Dittmaier$^2$, M. Roth$^{3}$ and
D.~Wackeroth$^4$} 

\address{$^1$ Paul Scherrer Institut, Villigen, Switzerland}

\address{$^2$ Theoretische Physik, Universit\"at Bielefeld, Germany}

\address{$^3$ Institut f\"ur Theoretische Physik, Universit\"at Leipzig,
Germany}

\address{$^4$ Department of Physics and Astronomy, 
University of Rochester, USA}

\begin{abstract}
{\sc RacoonWW} is an event generator
for $\Pep\Pem\to\PW\PW\to 4\,$fermions$(+\gamma)$ that includes full
tree-level predictions for $\Pep\Pem\to 4f$ and $\Pep\Pem\to
4f+\gamma$ as well as ${\cal O}(\alpha)$ corrections to $\Pep\Pem\to4f$ 
in the so-called double-pole approximation. We briefly sketch the concept 
of the calculation on which this generator is based and present
some numerical results.
\end{abstract}




\section{Introduction}

At LEP2 and future $\Pep\Pem$ linear colliders, 
the most important processes to study the
properties of the W~boson are  $\eeWWffff$. 
For integrated quantities the accuracy typically
reaches the order of 1\% at LEP2 and will even exceed the per-cent level
at future colliders. To account for this precision in 
predictions is a non-trivial task.

High-precision calculations for four-fermion production are
complicated for various reasons. At the aimed accuracy of some 0.1\%,
a pure on-shell approximation for the W~bosons is not acceptable,
i.e.\ the W~bosons have to be treated as resonances. Since the
description of resonances necessarily goes beyond a fixed-order
calculation in perturbation theory, problems with gauge invariance
occur.  Discussions of this issue can be found in
\citeres{lep2repWcs,flscheme}.  A second complication arises from the
need to take into account
electroweak radiative corrections 
beyond the universal corrections.  The full treatment of the processes
$\Pep\Pem\to 4f$ at the one-loop level is of enormous complexity and
involves severe theoretical problems with gauge invariance; up to now
such results have not been published.

Since lowest-order calculations for $\Pep\Pem\to 4f$ have already been
extensively discussed in the literature (see e.g.\ \citere{lep2repWevgen} 
and references therein), 
we concentrate on radiative corrections in the following.
Recent results of a full lowest-order calculation for real
photon emission, $\Pep\Pem\to 4f+\gamma$ are reviewed in \refse{se:ee4fg}.
In \refse{se:ee4f} we summarize the strategy and present some results
of the event generator {\sc RacoonWW}, which combines full tree-level
predictions for $\Pep\Pem\to 4f$ and $\Pep\Pem\to 4f+\gamma$ with an
approximation for the virtual ${\cal O}(\alpha)$ corrections to
four-fermion production via a resonant W-boson pair,
$\Pep\Pem\to\PW\PW\to 4f$.

\section{Full tree-level predictions for \boldmath{$\Pep\Pem\to 4f+\gamma$}}
\label{se:ee4fg}

The processes $\Pep\Pem\to 4f+\gamma$ are interesting mainly for two
reasons. On the one hand, they are a source of (real) ${\cal O}(\alpha)$ 
corrections to gauge-boson pair production with a four-fermion final 
state. On the other hand, they are 
sensitive to anomalous quartic gauge-boson couplings, such as 
$\gamma\gamma\PW\PW$, $\gamma\PZ\PW\PW$, and $\gamma\gamma\PZ\PZ$.
In the following we briefly summarize some results of \citere{ee4fa},
where the building block of the event generator {\sc RacoonWW}
\cite{racoonww1,racoonww2} is described that calculates cross sections for
$\Pep\Pem\to 4f+\gamma$ with arbitrary massless fermions.

In this event generator 
different schemes for treating gauge-boson widths are implemented. 
A comparison of results obtained in these different schemes
is useful in order to get information about the size of
gauge-invariance-breaking effects, which are present in some
finite-width schemes.
\begin{table}
\begin{center}
{\begin{tabular}{|c|c|r@{}l|r@{}l|r@{}l|r@{}l|}
\hline
\multicolumn{1}{|c|}{$\si/\fb$} &
\multicolumn{1}{r|}{$\sqrt{s}=$} & 
\multicolumn{2}{c|}{$189\GeV$} & 
\multicolumn{2}{c|}{$500\GeV$} &
\multicolumn{2}{c|}{$2\TeV$} & 
\multicolumn{2}{c|}{$10\TeV$}
\\\hline\hline
& constant width
&$    224.0 $&$( 4)$
&$     83.4 $&$( 3)$
&$     6.98 $&$( 5)$
&$    0.457 $&$( 6)$
\\\cline{2-10}
$\Pep \Pem \to \Pu\, \Pdbar\, \mu^- \bar{\nu}_\mu \,\ga$
& running width
&$    224.6 $&$( 4)$
&$     84.2 $&$( 3)$
&$     19.2 $&$( 1)$
&$      368 $&$( 6)$
\\\cline{2-10}
& complex mass          
&$    223.9 $&$( 4)$
&$     83.3 $&$( 3)$
&$     6.98 $&$( 5)$
&$    0.460 $&$( 6)$
\\\hline\hline
& constant width
&$    230.0 $&$( 4)$
&$    136.5 $&$( 5)$
&$     84.0 $&$( 7)$
&$     16.8 $&$( 5)$
\\\cline{2-10}
$\Pep \Pem \to \Pu\, \Pdbar\, \Pe^- \Pnebar \,\ga $
& running width
&$    230.6 $&$( 4)$
&$    137.3 $&$( 5)$
&$     95.7 $&$( 7)$
&$      379 $&$( 6)$
\\\cline{2-10}
& complex mass          
&$    229.9 $&$( 4)$
&$    136.4 $&$( 5)$
&$     84.1 $&$( 6)$
&$     16.8 $&$( 5)$
\\\hline
\end{tabular}}
\end{center}
\caption[]{Comparison of different finite-width schemes 
(taken from \citere{ee4fa})}
\label{tab:ee4fawidth}
\end{table}
Table~\ref{tab:ee4fawidth} contains some 
results on the total cross section for 
two semi-leptonic four-fermion final states and a photon, evaluated 
with different finite-width treatments.
Similar to the case without photon emission, the SU(2)-breaking effects
induced by a running width render the predictions totally wrong in the
TeV range. For a constant width such effects are suppressed, as can be
seen from a comparison with the results of the complex-mass scheme,
which exactly preserves gauge invariance.

\begin{figure}
\centerline{
\setlength{\unitlength}{1cm}
\begin{picture}(6,6.6)
\put(0,0){\includegraphics{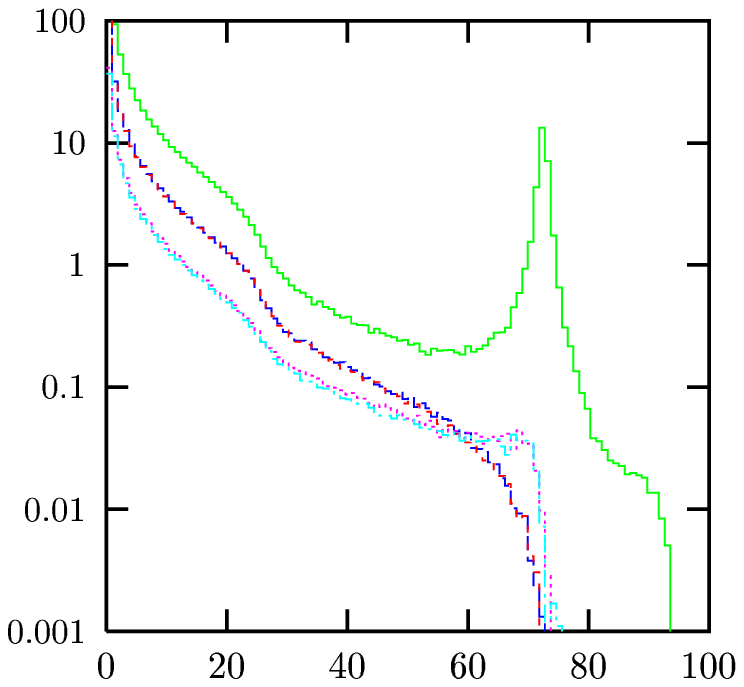}}
\put(1.5,2.1){\footnotesize $\sqrt{s}=189\GeV$}
\put(3.0,-0.1){\makebox(1,1)[cc]{{\small $E_\ga/\GeV$}}}
\end{picture}
\begin{picture}(6,6.6)
\put(0,0){\includegraphics{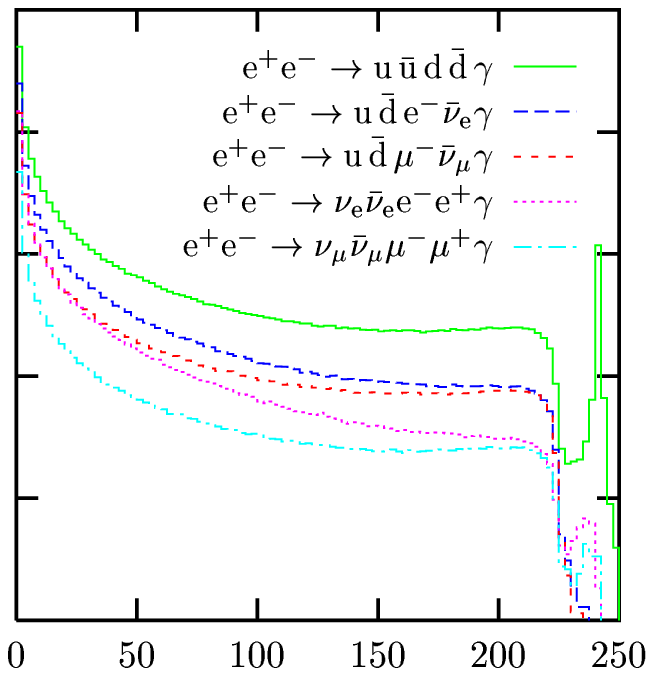}}
\put(1.1,2.1){\footnotesize $\sqrt{s}=500\GeV$}
\put(2.5,-0.1){\makebox(1,1)[cc]{{\small $E_\ga/\GeV$}}}
\end{picture} }
\caption[]{Photon-energy spectra 
$(\rd\si/\rd E_\ga)/(\fb/\GeV)$ for several processes 
(taken from \citere{ee4fa})}
\label{fi:photonspectra}
\end{figure}
\begin{figure}
\centerline{
\begin{picture}(150,110)(-5,0)
\SetScale{.8}
\ArrowLine(0,10)(40,50)
\ArrowLine(40,70)(0,110)
\Photon(50,60)(150,60){2}{11}
\Photon(50,63)(120,90){2}{8}
\Photon(50,57)(120,30){2}{8}
\Vertex(120,90){2}
\Vertex(120,30){2}
\ArrowLine(120,90)(150,110)
\ArrowLine(150,70)(120,90)
\ArrowLine(120,30)(150,50)
\ArrowLine(150,10)(120,30)
\GCirc(50,60){15}{0}
\put(75,18){\makebox(1,1)[c]{$V_2$}}
\put(75,78){\makebox(1,1)[c]{$V_1$}}
\put(128,50){\makebox(1,1)[c]{$\gamma$}}
\Text(-5,100)[rt]{a)}
\SetScale{1}
\end{picture}
\begin{picture}(160,100)
\SetScale{.8}
\ArrowLine(0,10)(40,50)
\ArrowLine(40,70)(0,110)
\Photon(50,63)(150,100){2}{12}
\Photon(50,57)(110,30){2}{8}
\Vertex(110,30){2}
\ArrowLine(110,30)(140,50)\ArrowLine(140,50)(170,70)
\ArrowLine(150,10)(110,30)
\Vertex(140,50){2}
\Photon(140,50)(170,30){2}{4}
\Vertex(170,30){2}
\ArrowLine(170,30)(200,50)
\ArrowLine(200,10)(170,30)
\GCirc(50,60){15}{0}
\put(65,24){\makebox(1,1)[c]{$Z$}}
\put(133,38){\makebox(1,1)[c]{$V_3$}}
\put(128,80){\makebox(1,1)[c]{$\gamma$}}
\Text(-5,100)[rt]{b)}
\SetScale{1}
\end{picture}
}
\caption[]{Diagrams for important subprocesses in $4f+\gamma$ production
($V_1,V_2=\PW,\PZ,\ga$, $V_3=\ga,\Pg$) }
\label{fi:resonance}
\end{figure}
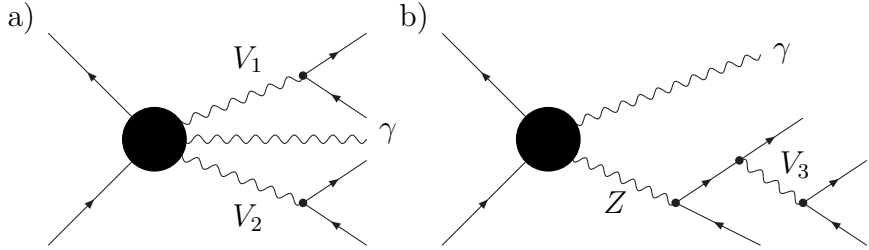

Figure~\ref{fi:photonspectra} shows the photon-energy spectra for some
typical four-fermion final states that correspond to $\PW\PW\gamma$
production. Apart from the usual soft-photon pole, the spectra contain
several threshold and peaking structures that are caused by photon
emission from the initial state. The two relevant classes of diagrams
are illustrated in \reffi{fi:resonance}.
Diagrams with
the structure of \reffi{fi:resonance}a correspond to
triple-gauge-boson-production subprocesses and yield dominant
contributions as long as the two virtual gauge bosons can become
simultaneously resonant. For instance, $\PW\PW\gamma$ production is
dominant for $E_\gamma < 26.3\GeV$ ($224\GeV$) for a centre-of-mass (CM)
energy of $189\GeV$ ($500\GeV$). The diagrams of \reffi{fi:resonance}b
correspond to $\gamma\PZ$ production with a subsequent four-particle
decay of the resonant Z~boson mediated by a soft photon or gluon $V_3$.
Owing to the two-particle kinematics of $\gamma\PZ$ production such
contributions lead to peaking
structures around a fixed value of $E_\gamma$,
which is located at $72.5\GeV$ ($242\GeV$) for a CM energy of $189\GeV$
($500\GeV$). 

Table \ref{tab:ee4fawidth} and \reffi{fi:photonspectra} also illustrate the
effect of background diagrams, since final states that are related by
the interchange of muons and electrons differ only by background
diagrams. 
While the impact of background diagrams is of the order of some per cent
for CM energies around $200\GeV$, there is a large effect of background 
contributions already at $500\GeV$.
The main effect is due to forward-scattered $\Pe^\pm$, which is familiar
from the results on $\Pep\Pem\to 4f$. More numerical results for
$\Pep\Pem\to 4f+\gamma$ can be found in \citeres{ee4fa,ee4fa0}.

\section{Electroweak radiative corrections to 
\boldmath{$\Pep\Pem\to\PW\PW\to 4f$}}
\label{se:ee4f}

\subsection{Relevance of electroweak corrections}

In the past, Monte Carlo generators
for off-shell W-pair production (see e.g.\ \citere{lep2repWevgen})
typically included only universal electroweak ${\cal O}(\alpha)$ corrections
such as the running of the electromagnetic coupling, $\alpha(q^2)$, 
leading corrections entering via the $\rho$-parameter,
the Coulomb singularity, which is important near threshold,
and mass-singular logarithms 
$\alpha\ln(\Me^2/Q^2)$ from initial-state radiation, where $Q^2$ is not
determined and has to be set to a typical scale for the process under
consideration.
The size of the remaining ${\cal O}(\alpha)$ contributions is
estimated by inspecting on-shell W-pair production, for which the exact 
${\cal O}(\alpha)$ correction and the leading contributions were
given in \citeres{vrceeww} and \cite{bo92}, respectively.
The difference $\de_{\mathrm{IBA}}-\de$ between an ``improved
Born approximation'' $\de_{\mathrm{IBA}}$, which is based on the 
above-mentioned universal corrections, and the corresponding full
${\cal O}(\alpha)$ correction $\de$ corresponds to the
non-leading corrections and has
already been discussed in \citeres{lep2repWcs,crad96}.
For total cross sections
this difference amounts to $\sim 1$--2\% for LEP2 energies,
but to $\sim 10$--20\% in the TeV range; for distributions the
difference is even larger in general. Thus, in view of a desired
accuracy of some $0.1\%$, the inclusion of non-leading corrections is
indispensable. 

\subsection{Electroweak corrections in double-pole approximation}

Fortunately, the full off-shell calculation for the processes
$\Pep\Pem\to\PW\PW\to 4f$ in ${\cal O}(\alpha)$ is not needed for
most applications. Sufficiently far above the W-pair threshold 
a good approximation can be obtained by taking into account only those
contributions that are enhanced by two resonant \PW~bosons. The uncertainty
from neglecting corrections to background diagrams can be estimated
to some $0.1\%$, at least in the absence of special enhancement effects
such as forward-scattered $\Pe^\pm$.
Doubly-resonant corrections to 
$\Pep\Pem\to\PW\PW\to 4f$ can be classified into two types
\cite{lep2repWcs,ae94,be94}:
factorizable and non-factorizable corrections.

{\it Factorizable corrections} are those that correspond either 
to W-pair production or to W~decay. We first focus on virtual 
factorizable corrections, which are represented by the schematic diagram of 
\reffi{fig:fRCsdiag}, in which the shaded blobs contain all one-loop
corrections to the production and decay processes, and the open blobs
include the corrections to the W~propagators. 
\begin{figure}
\centerline{
\begin{picture}(155,95)(0,-10)
\SetScale{.8}
\ArrowLine(30,50)( 5, 95)
\ArrowLine( 5, 5)(30, 50)
\Photon(30,50)(150,80){2}{11}
\Photon(30,50)(150,20){2}{11}
\ArrowLine(150,80)(190, 95)
\ArrowLine(190,65)(150,80)
\ArrowLine(190, 5)(150,20)
\ArrowLine(150,20)(190,35)
\GCirc(30,50){10}{.5}
\GCirc(90,65){10}{1}
\GCirc(90,35){10}{1}
\GCirc(150,80){10}{.5}
\GCirc(150,20){10}{.5}
\DashLine( 70,0)( 70,100){2}
\DashLine(110,0)(110,100){2}
\put(40,21){W}
\put(40,53){W}
\put(95, 8){W}
\put(95,67){W}
\put(-12, 5){$\Pem$}
\put(-12,70){$\Pep$}
\put(160, 1){$\bar f_4$}
\put(160,24){$f_3$}
\put(160,50){$\bar f_2$}
\put(160,75){$f_1$}
\put(-25,-10){\footnotesize On-shell production}
\put(100,-10){\footnotesize On-shell decays}
\SetScale{1}
\end{picture}
}
\caption{Diagrammatic structure of virtual factorizable corrections to
$\Pep\Pem\to\PW\PW\to 4f$}
\label{fig:fRCsdiag}
\end{figure}
The corresponding matrix element is of the form
\beqar
\label{eq:Mstruc}
{\cal M} =
\underbrace{\frac{R_{+-}(k_+^2,k_-^2)}{(k_+^2-\MW^2)(k_-^2-\MW^2)}
        }_{\mbox{doubly-resonant}}
+\underbrace{\frac{R_{+}(k_+^2,k_-^2)}{k_+^2-\MW^2}
            +\frac{R_{-}(k_+^2,k_-^2)}{k_-^2-\MW^2}
        }_{\mbox{singly-resonant}}
+\underbrace{N(k_+^2,k_-^2)}_{\mbox{non-resonant}}\!\!\!,
\nn\\
\eeqar
and the double-pole approximation (DPA) amounts to the replacement
\beq
{\cal M} \to
\frac{R_{+-}(\MW^2,\MW^2)}{(k_+^2-\MW^2+\ri\MW\GW)(k_-^2-\MW^2+\ri\MW\GW)},
\eeq
where the originally gauge-dependent numerator $R_{+-}(k_+^2,k_-^2)$ 
is replaced by the gauge-independent residue $R_{+-}(\MW^2,\MW^2)$ 
\cite{st91,ae94}.
The one-loop corrections to this residue can be deduced from the known
results for the pair production \cite{vrceeww}
and the decay \cite{rcwdecay} of on-shell W~bosons.

The formulation of a consistent DPA for the real
corrections, and thus a splitting into factorizable and non-factorizable
parts, is also possible, but non-trivial. The main complication originates 
from the emission of photons from the resonant W~bosons. 
A diagram with a radiating W~boson involves two propagators 
with momenta that differ only by the
momentum of the emitted photon. If the photon momentum is large 
($E_\gamma\gg\Gamma_\PW$), the resonances of these two propagators are
well separated in phase space, and their contributions can be associated
with photon radiation from exactly one of the production or decay
subprocesses. For soft photons ($E_\gamma\ll\Gamma_\PW$),
the resonances coincide, and the DPA is identical to the one without photon.
However, for $E_\gamma\sim\Gamma_\PW$ the two
resonance factors for the radiating W~boson overlap.
Although it is possible to decompose these factors into contributions
associated with the subprocesses, a reliable estimate of the accuracy of 
the corresponding DPA is not obvious.

{\it Non-factorizable corrections} comprise all those doubly-resonant
corrections that are not yet contained in the factorizable ones, i.e.\ 
they include all diagrams involving particle exchange between the
subprocesses.  These corrections do not contain the product of two
independent Breit--Wigner-type resonances for the W~bosons, i.e.\ the
production and decay subprocesses are not independent in this case.
Simple power-counting arguments reveal that such diagrams only lead to
doubly-resonant contributions if the exchanged particle is a photon
with energy $E_\gamma\lsim\Gamma_\PW$; all other non-factorizable
diagrams are negligible in DPA. Two relevant diagrams are shown in
\reffi{fig:nfRCsdiags}, where the full blobs represent tree-level
subgraphs.
\begin{figure}
\centerline{
\begin{picture}(110,75)(0,8)
\SetScale{0.8}
\ArrowLine(30,50)( 5, 95)
\ArrowLine( 5, 5)(30, 50)
\Photon(30,50)(90,80){2}{6}
\Photon(30,50)(90,20){2}{6}
\GCirc(30,50){10}{0}
\Vertex(90,80){1.2}
\Vertex(90,20){1.2}
\ArrowLine(90,80)(120, 95)
\ArrowLine(120,65)(105,72.5)
\ArrowLine(105,72.5)(90,80)
\Vertex(105,72.5){1.2}
\ArrowLine(120, 5)( 90,20)
\ArrowLine( 90,20)(105,27.5)
\ArrowLine(105,27.5)(120,35)
\Vertex(105,27.5){1.2}
\Photon(105,27.5)(105,72.5){2}{4.5}
\put(89,40){$\gamma$}
\put(42,60){$W$}
\put(42,15){$W$}
\SetScale{1}
\end{picture}
\begin{picture}(190,75)(0,8)
\SetScale{.8}
\ArrowLine(30,50)( 5, 95)
\ArrowLine( 5, 5)(30, 50)
\Photon(30,50)(90,80){2}{6}
\Photon(30,50)(90,20){2}{6}
\GCirc(30,50){10}{0}
\Vertex(90,80){1.2}
\Vertex(90,20){1.2}
\ArrowLine(90,80)(120, 95)
\ArrowLine(120,65)(105,72.5)
\ArrowLine(105,72.5)(90,80)
\ArrowLine(120, 5)( 90,20)
\ArrowLine( 90,20)(120,35)
\Vertex(105,72.5){1.2}
\PhotonArc(120,65)(15,150,270){2}{3}
\put(42,60){W}
\put(42,15){W}
\put(75,40){$\gamma$}
\DashLine(120,0)(120,100){6}
\PhotonArc(120,35)(15,-30,90){2}{3}
\Vertex(135,27.5){1.2}
\ArrowLine(150,80)(120,95)
\ArrowLine(120,65)(150,80)
\ArrowLine(120, 5)(150,20)
\ArrowLine(150,20)(135,27.5)
\ArrowLine(135,27.5)(120,35)
\Vertex(150,80){1.2}
\Vertex(150,20){1.2}
\Photon(210,50)(150,80){2}{6}
\Photon(210,50)(150,20){2}{6}
\ArrowLine(210,50)(235,95)
\ArrowLine(235, 5)(210,50)
\GCirc(210,50){10}{0}
\put(140,60){W}
\put(140,15){W}
\SetScale{1}
\end{picture}
}
\caption{Examples of virtual and real non-factorizable 
corrections to $\Pep\Pem\to\PW\PW\to 4f$}
\label{fig:nfRCsdiags}
\end{figure}
We note that diagrams involving photon exchange between the W~bosons
contribute both to factorizable and
non-factorizable corrections; otherwise the splitting into those parts
would not be gauge-invariant.
The calculation of non-factorizable corrections to 
$\Pep\Pem\to\PW\PW\to 4f$ was discussed in \citeres{me96,be97,de98a} 
in detail. A numerical discussion of the sum of virtual and real
non-factorizable corrections can be found in \citeres{be97,de98a}.

\subsection{Results for ${\cal O}(\alpha)$ corrections in double-pole
approximation}

Different versions of DPA have already been used in the literature
\cite{be98b,ja97,ja99,ku99}. For instance,
following a semi-analytical approach,
a consistent application of the DPA for the virtual and real corrections
to four-lepton production was presented in \citere{be98b}.
In such an approach, however, it
is not possible to apply all experimentally relevant phase-space cuts,
and effects of recombining photons with nearly collinear charged
fermions cannot be treated realistically.
For the latter reason, in \citere{be98b} 
the invariant masses of the \PW~bosons were defined strictly in terms
of the invariant masses of the corresponding decay fermion pairs.
As a result, the invariant-mass distributions in
$M_\pm=\sqrt{k_\pm^2}$ received large corrections from final-state
radiation, namely $-20\MeV$, $-39\MeV$, and $-77\MeV$ for
$\tau^+\nu_\tau$, $\mu^+\nu_\mu$, and $\Pep\nu_\Pe$ final states at
$\sqrt{s}=184\GeV$, respectively.  These results have been
qualitatively confirmed by YFSWW in \citere{ja99}, where the 
${\cal O}(\alpha)$ corrections to W-pair production \cite{ja97} were
supplemented by final-state radiation in a leading-log approach.
Note, however, that the large shifts are due to mass-singular
logarithms like $\alpha\ln(m_l/\MW)$, 
which occur because no recombination of the fermions with collinear
photons is performed.
More realistic definitions of
$k_\pm^2$, which have to include photon recombination, effectively
replace the mass-singular logarithms by logarithms of a minimum
opening angle for collinear photon emission. This expectation was also
confirmed in \citere{ja99}.

At present, {\sc RacoonWW} \cite{racoonww1,racoonww2} is the only Monte Carlo
event generator with a complete implementation of the ${\cal O}(\alpha)$
corrections to $\Pep\Pem\to\PW\PW\to 4f$ in DPA. In RacoonWW, only
the virtual corrections are treated in DPA, whereas the real-photonic
corrections are based on the full lowest-order calculation for
$\Pep\Pem\to 4f+\gamma$ described in \citere{ee4fa}. This means, in
particular, that only the virtual corrections are split into
factorizable and non-factorizable corrections. Note that this approach
requires a careful treatment of IR and mass singularities, since the
singularity structure in the virtual and real corrections are related to
different lowest-order cross sections. For the virtual corrections the
DPA Born cross section for $\Pep\Pem\to\PW\PW\to 4f$ is relevant, while
for the real corrections the full Born cross section  for $\Pep\Pem\to 4f$
applies. In {\sc RacoonWW} one can choose between two different methods
for the matching between virtual and real corrections; one method is
based on phase-space slicing, the other on the subtraction method
described in \citere{subtract}. More details about the {\sc RacoonWW}
approach can be found in \citeres{racoonww1,racoonww2}.

\begin{figure}
\centerline{%
\setlength{\unitlength}{1cm}
\begin{picture}(7.9,7.2)
\put(0,0){\includegraphics{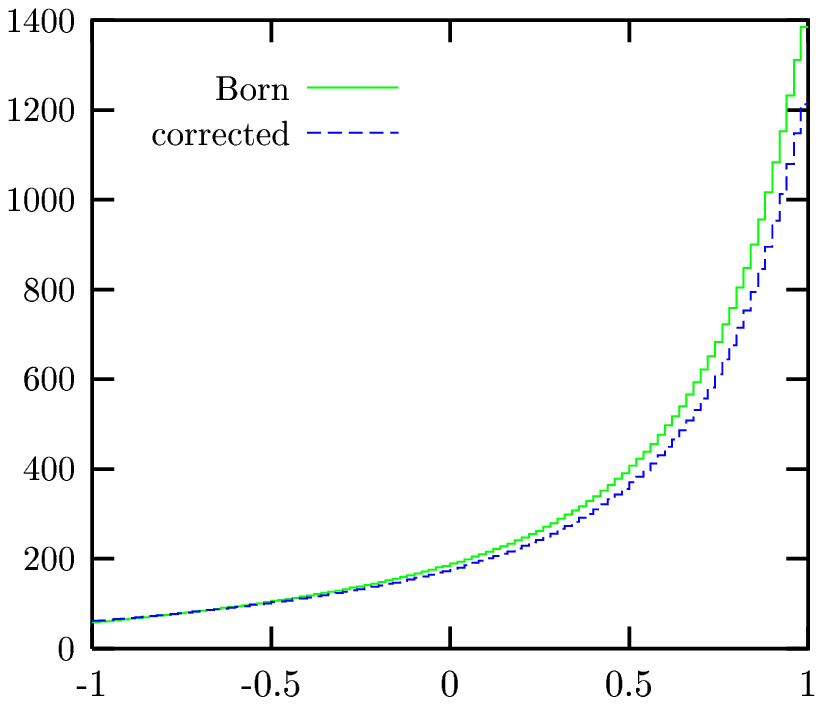}}
\put(4.0,-0.2){\makebox(1,1)[c]{$\cos\theta_{\PW}$}}
\end{picture}%
\begin{picture}(7.9,7.2)
\put(4.0,-0.2){\makebox(1,1)[c]{$\cos\theta_{\PW}$}}
\put(0,0){\includegraphics{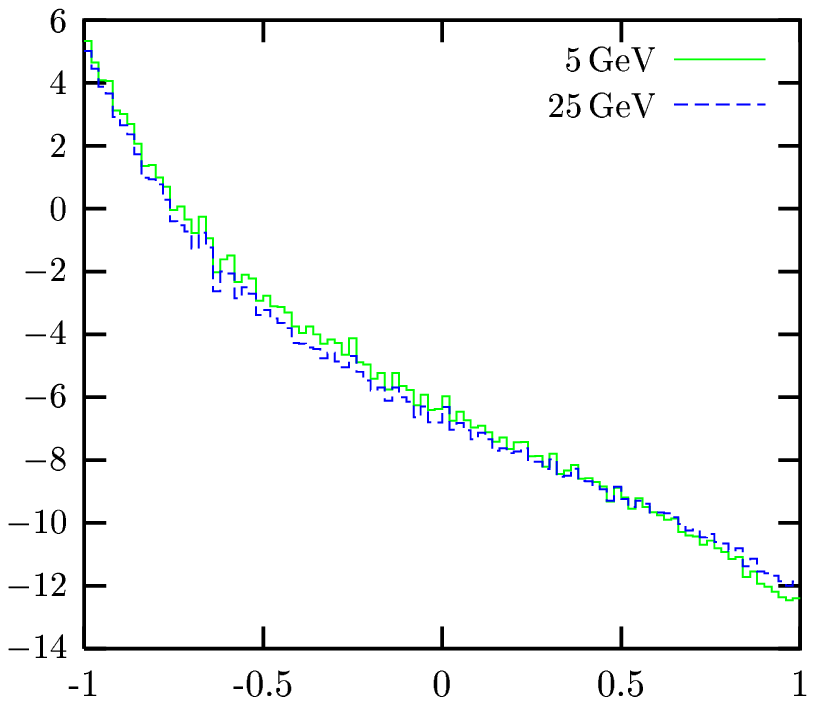}}
\end{picture}
}
\vspace*{-.3em}
\caption{Production-angle distribution 
$(\rd \si/\rd \cos\theta_{\PW})/\fb$ (left) and relative correction
$\de/\%$ (right) for $\Pep\Pem\to\nu_\mu\mu^+\Pd\Pubar$ at 
$\protect\sqrt{s}=200\GeV$ (taken from \citere{racoonww1})}
\label{fi:prod_angle}
\vspace*{1em}
\centerline{%
\setlength{\unitlength}{1cm}
\begin{picture}(7.9,7.2)
\put(0,0){\includegraphics{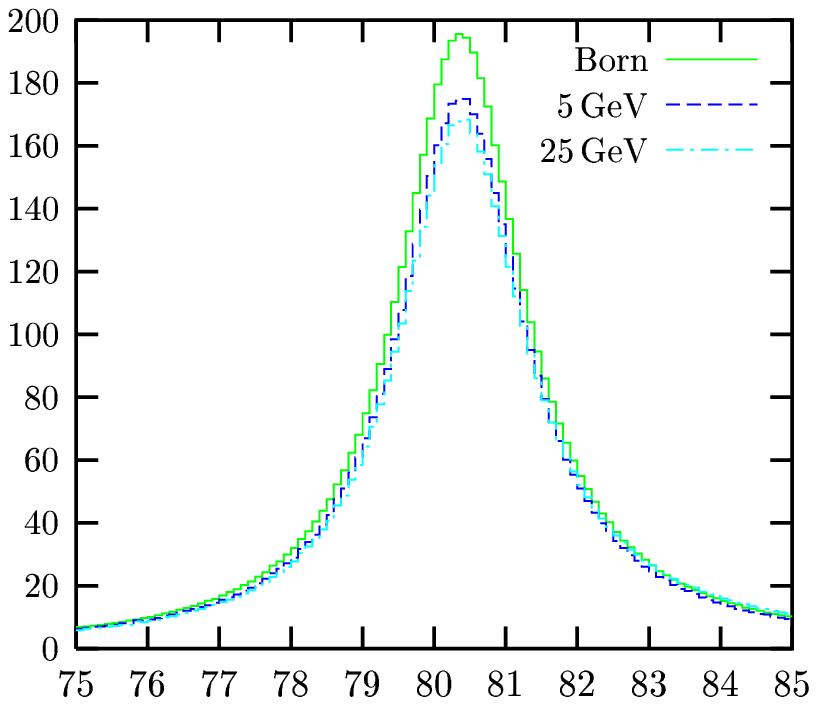}}
\put(4.0,-0.2){\makebox(1,1)[c]{$M_{\Pd\Pu}\ [\mathrm{GeV}]$}}
\end{picture}%
\begin{picture}(7.9,7.2)
\put(4.0,-0.2){\makebox(1,1)[c]{$M_{\Pd\Pu}\ [\mathrm{GeV}]$}}
\put(0,0){\includegraphics{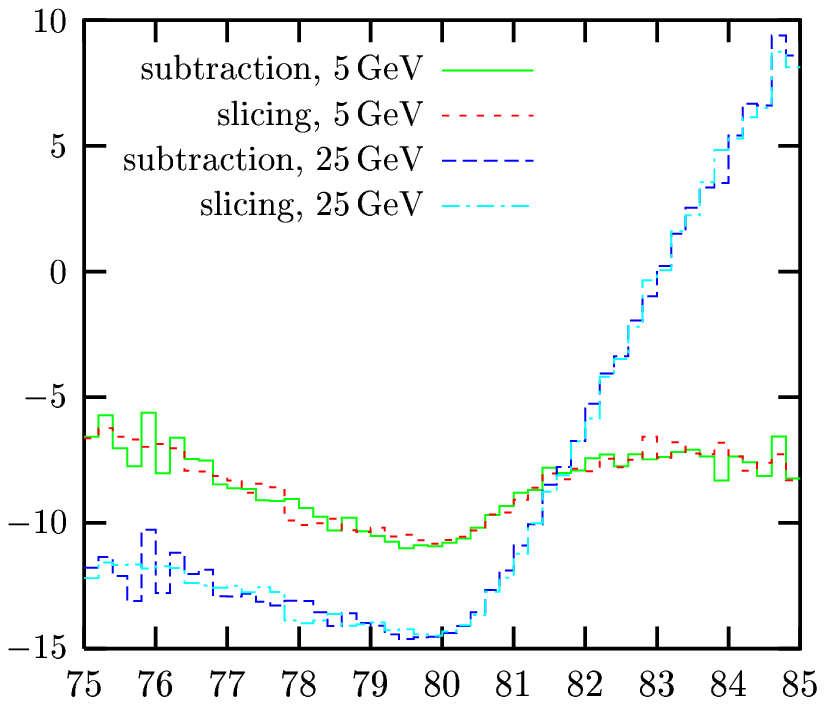}}
\end{picture}
}
\vspace*{-.3em}
\caption{Invariant-mass distribution 
$(\rd \si/\rd M_{\Pd\Pu})/(\fb/\GeV)$ of the quark pair (left) and 
relative correction $\de/\%$ (right) for
$\Pep\Pem\to\nu_\mu\mu^+\Pd\Pubar$ at $\protect\sqrt{s}=200\GeV$
(taken from \citere{racoonww1})}
\label{fi:ud_invmass}
\end{figure}%
Figures~\ref{fi:prod_angle} and \ref{fi:ud_invmass} show some results
of {\sc RacoonWW} for the semileptonic reaction
$\Pep\Pem\to\nu_\mu\mu^+\Pd\Pubar$ at the typical LEP2 CM energy $200\GeV$,
where $\theta_\PW$ is the W~production angle (with the W~momentum
being defined by the total
momentum of $\nu_\mu$ and $\mu^+$) and $M_{\Pd\Pu}$ is the invariant mass of
the quark pair. The precise definition of the input and of the
photon-recombination procedure, as well as a detailed discussion of more
results, can be found in \citere{racoonww1}. 
For the photon recombination we first determine the lowest invariant
mass $M_{\gamma f}$ built by an emitted photon and a charged final-state 
fermion. If $M_{\gamma f}$ is smaller than a fixed recombination cut
$M_{\mathrm{rec}}$ the photon momentum is added to the one of the 
corresponding fermion $f$.
The curves denoted by ``$5\GeV$'' and ``$25\GeV$'' correspond to the
respective values of $M_{\mathrm{rec}}$.
While the production-angle
distribution is not very sensitive to $M_{\mathrm{rec}}$, 
the invariant-mass distribution strongly 
depends on the recombination
procedure, as expected from the above discussion.
For different recombination procedures the maxima of the line shapes
differ by up to \mbox{30--40}$\MeV$ \cite{racoonww1}. As can be seen from 
\reffi{fi:ud_invmass}, there is a tendency to shift the maxima to larger
invariant masses if more and more photons are recombined.

\section*{References}

\frenchspacing
\newcommand{\app}[3]{{\sl Acta Phys.\ Pol.} {\bf #1} (19#2) #3}
\newcommand{\ap}[3]{{\sl Ann.~Phys.} {\bf #1} (19#2) #3}
\newcommand{\zp}[3]{{\sl Z.~Phys.} {\bf #1} (19#2) #3}
\newcommand{\np}[3]{{\sl Nucl.~Phys.} {\bf #1} (19#2) #3}
\newcommand{\pl}[3]{{\sl Phys.~Lett.} {\bf #1} (19#2) #3}
\newcommand{\prep}[3]{{\sl Phys.\ Rep.} {\bf #1} (19#2) #3}
\newcommand{\pr}[3]{{\sl Phys.~Rev.} {\bf #1} (19#2) #3}
\newcommand{\prl}[3]{{\sl Phys.~Rev.~Lett.} {\bf #1} (19#2) #3}
\newcommand{\fp}[3]{{\sl Fortschr.~Phys.} {\bf #1} (19#2) #3}
\newcommand{\jp}[3]{{\sl J.~Phys.} {\bf #1} (19#2) #3}
\newcommand{\cpc}[3]{{\sl Comput.~Phys.~Commun.} {\bf #1} (19#2) #3}
\newcommand{\ijmp}[3]{{\sl Int.~J.~Mod.~Phys.} {\bf #1} (19#2) #3}
\newcommand{\nim}[3]{{\sl Nucl.~Instr.~Meth.} {\bf #1} (19#2) #3}
\newcommand{\nc}[3]{{\sl Nuovo Cimento} {\bf #1} (19#2) #3}
\newcommand{\vj}[4]{{\sl #1} {\bf #2} (19#3) #4}

\end{document}